\documentclass[epj]{svjour}
\usepackage{graphics}
\usepackage{xcolor}

\usepackage{wrapfig,lipsum,booktabs}
\usepackage{color,graphicx}
\usepackage{amsmath}
\usepackage{amssymb}
\usepackage[mathlines]{lineno}
\usepackage{mathtools}
\usepackage[hidelinks]{hyperref}  %

\begin{document}
\title{Towards an automated data cleaning with deep learning in CRESST}
%

\author{
  G.~Angloher \inst{1}\and
  S.~Banik\inst{2,3}\and
  D.~Bartolot\inst{2}\and
  G.~Benato\inst{4}\and
  A.~Bento\inst{1,9}\and 
  A.~Bertolini\inst{1}\and 
  R.~Breier\inst{5}\and
  C.~Bucci\inst{4}\and 
  J.~Burkhart\inst{2}\and
  L.~Canonica\inst{1}\and 
  A.~D'Addabbo\inst{4}\and
  S.~Di~Lorenzo\inst{4}\and
  L.~Einfalt\inst{2,3}\and
  A.~Erb\inst{6,10}\and
  F.~v.~Feilitzsch\inst{6}\and 
  N.~Ferreiro~Iachellini\inst{1}\and
  S.~Fichtinger\inst{2}\and
  D.~Fuchs\inst{1}\and 
  A.~Fuss\inst{2,3}\and
  A.~Garai\inst{1}\and 
  V.M.~Ghete\inst{2}\and
  S.~Gerster\inst{7}\and
  P.~Gorla\inst{4}\and
  P.V.~Guillaumon\inst{4}\and
  S.~Gupta\inst{2}\and 
  D.~Hauff\inst{1}\and 
  M.~Ješkovsk\'y\inst{5}\and
  J.~Jochum\inst{7}\and
  M.~Kaznacheeva\inst{6}\and
  A.~Kinast\inst{6}\and
  H.~Kluck\inst{2}\and
  H.~Kraus\inst{8}\and 
  M.~Lackner\inst{1}\and
  A.~Langenk\"amper\inst{1,6}\and 
  M.~Mancuso\inst{1}\and
  L.~Marini\inst{4,11}\and 
  L.~Meyer\inst{7}\and
  V.~Mokina\inst{2}\and
  A.~Nilima\inst{1}\and 
  M.~Olmi\inst{4}\and
  T.~Ortmann\inst{6}\and
  C.~Pagliarone\inst{4,12}\and
  L.~Pattavina\inst{4,6}\and
  F.~Petricca\inst{1}\and 
  W.~Potzel\inst{6}\and 
  P.~Povinec\inst{5}\and
  F.~Pr\"obst\inst{1}\and
  F.~Pucci\inst{1}\and 
  F.~Reindl\inst{2,3} \and
  D.~Rizvanovic\inst{2}\and
  J.~Rothe\inst{6}\and 
  K.~Sch\"affner\inst{1}\and 
  J.~Schieck\inst{2,3}\and 
  D.~Schmiedmayer\inst{2,3}\and
  S.~Sch\"onert\inst{6}\and 
  C.~Schwertner\inst{2,3}\and
  M.~Stahlberg\inst{1}\and 
  L.~Stodolsky\inst{1}\and 
  C.~Strandhagen\inst{7}\and
  R.~Strauss\inst{6}\and
  I.~Usherov\inst{7}\and
  F.~Wagner\inst{2}\thanks{felix.wagner@oeaw.ac.at}\and 
  M.~Willers\inst{6}\and 
  V.~Zema\inst{1}
(CRESST Collaboration)
\and 
W.~Waltenberger\inst{2}
}

\institute
{Max-Planck-Institut f\"ur Physik, D-80805 M\"unchen, Germany \label{addrMPI} 
\and
Institut f\"ur Hochenergiephysik der \"Osterreichischen Akademie der Wissenschaften, A-1050 Wien, Austria\label{addrHEPHY} 
\and
Atominstitut, Technische Universit\"at Wien, A-1020 Wien, Austria \label{addrAI} 
\and
INFN, Laboratori Nazionali del Gran Sasso, I-67100 Assergi, Italy \label{addrLNGS} 
\and
Comenius University, Faculty of Mathematics, Physics and Informatics, 84248 Bratislava, Slovakia \label{addrBratislava} 
\and
Physik-Department, Technische Universit\"at M\"unchen, D-85747 Garching, Germany \label{addrTUM} 
\and
Eberhard-Karls-Universit\"at T\"ubingen, D-72076 T\"ubingen, Germany \label{addrTUE} 
\and
Department of Physics, University of Oxford, Oxford OX1 3RH, United Kingdom \label{addrOxford} 
\and
also at: LIBPhys-UC, Departamento de Fisica, Universidade de Coimbra, P3004 516 Coimbra, Portugal \label{addrCoimbra} 
\and
also at: Walther-Mei\ss ner-Institut f\"ur Tieftemperaturforschung, D-85748 Garching, Germany \label{addrWMI} 
\and
also at: GSSI-Gran Sasso Science Institute, I-67100 L'Aquila, Italy \label{addrGSSI} 
\and
also at: Dipartimento di Ingegneria Civile e Meccanica, Universitá degli Studi di Cassino e del Lazio Meridionale, I-03043 Cassino, Italy\label{addrCASS}
}

\date{Received: date / Revised version: date}
%

\abstract{
The CRESST experiment employs cryogenic calorimeters for the sensitive measurement of nuclear recoils induced by dark matter particles. The recorded signals need to undergo a careful cleaning process to avoid wrongly reconstructed recoil energies caused by pile-up and read-out artefacts. We frame this process as a time series classification task and propose to automate it with neural networks. With a data set of over one million labeled records from 68 detectors, recorded between 2013 and 2019 by CRESST, we test the capability of four commonly used neural network architectures to learn the data cleaning task. Our best performing model achieves a balanced accuracy of 0.932 on our test set. We show on an exemplary detector that about half of the wrongly predicted events are in fact wrongly labeled events, and a large share of the remaining ones have a context-dependent ground truth. We furthermore evaluate the recall and selectivity of our classifiers with simulated data. The results confirm that the trained classifiers are well suited for the data cleaning task. 
\PACS{
      {PACS-key}{discribing text of that key}   \and
      {PACS-key}{discribing text of that key}
     } 
} 

\maketitle
%


\section{Introduction}
\label{intro}

    
Dark Matter (DM) particles are hypothetical particles beyond the Standard Model of particle physics (SM) and thought to make up $(83.9 \pm 1.5 )$\% of all matter in our universe \cite{refId0}. The experimental search for particle DM inspired many experiments in the past decades. The Cryogenic Rare Event Search with Superconducting Thermometers (CRESST), located in the Laboratori Nazionali del Gran Sasso, is a direct DM search and uses scintillating, cryogenic calorimeters as targets. This technology achieves a) low nuclear recoil thresholds, with a currently lowest reported value of 10 eV \cite{https://doi.org/10.48550/arxiv.2207.09375}; b) currently the strongest exclusion limits on spin-independent (spin-dependent) DM-SM interactions for DM masses in the range 0.16-1.5 (0.25-1.5) GeV/c$^2$, under standard assumptions \cite{PhysRevD.100.102002,CRESSTLithiumLimit}. However, the sensitivity of the detectors and readout electronics cause not only particle recoils to trigger, but also a variety of artefacts: spikes, drifts, jumps and glitches in the noise baseline (BL) from the readout and heater electronics, and piled up pulse shapes (PSs) from multiple particle recoils in close temporal proximity. The recorded signals therefore need a cleaning step, before a meaningful data analysis can be started.


The standard approach is based on the calculation of PS and BL features, as e.g. the pulse height (PH), rise and decay time and BL slope. The cleaning is then carried out by an analyst, who defines individual rejection regions (cuts) in the space of the calculated features for each detector. Automating this process brings two benefits: a) detector setups with a large number of simultaneously operated detectors require less human effort to clean and analyse the recorded data and b) it helps preventing biases from individual decisions of the analyst made in the manual intervention. 

\begin{figure*}
 \includegraphics[width=\linewidth]{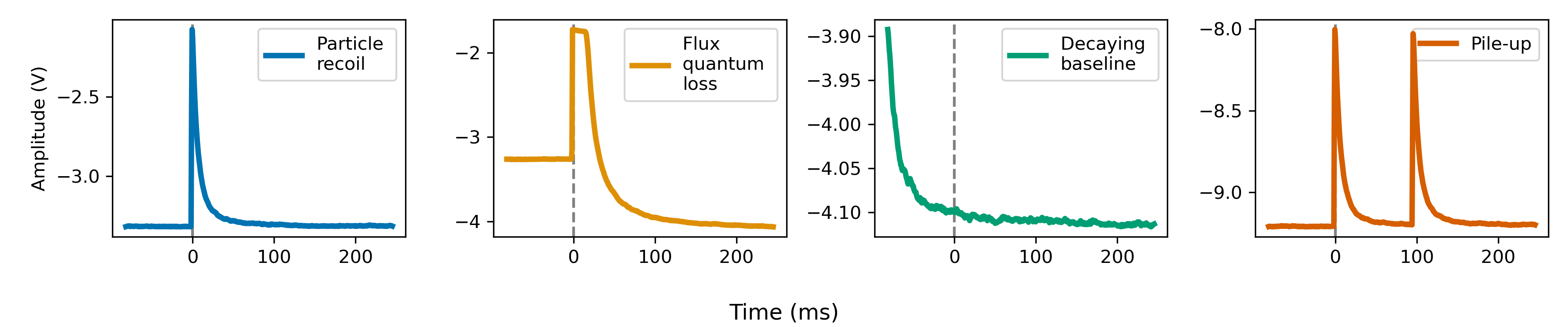}
\caption{Particle recoils produce a pulse-shaped record (blue). Flux quantum losses of the SQUID amplifier in the read-out circuit are caused by fast magnetic field changes, e.g. from high energy recoils (orange). Decaying BLs are residuals from earlier high energy pulses (green). Pile-up originates from multiple particle recoils within the same record window (red).}
\label{fig:events}       
\end{figure*}

In each detector a particle recoil produces a characteristic PS, determined by the thermal properties of the target and sensor. Artefacts usually deviate from this characteristic PS (see Fig.~\ref{fig:events}). 
One common multi-purpose data cleaning method is to fit the numerical array of the characteristic PS to each record and reject all those, whose fit error exceed a certain value. This method requires a relatively high computational cost for each record, either the prior knowledge of the detector-specific PS or dedicated training data from the same detector to build a template, and manual interventions in tuning the cut values. Furthermore, the discrimination power of cuts is often limited by overlaps of the artefact and target recoil feature distributions, where only correlated cuts on multiple features could achieve an optimal cleaning of the data.


    

In our work, we create a data set with samples from measurements that were done in CRESST-II and CRESST-III. We clean the data for each detector individually, and we label each record as accepted (positive) or rejected (negative). The data cleaning task is now equivalent to a binary, supervised time series classification problem, i.e. learning to discriminate between positive and negative records. We approach this task with deep neural networks, with which promising results were obtained for time series classification tasks \cite{ismail_fawaz_deep_2019}.  

\begin{wraptable}{r}{5cm}
\begin{tabular}{lr}\\\toprule  
Data set & Nmbr. of records \\\midrule
Training & 930,368 \\  \midrule
Validation & 49,024 \\  \midrule
Test & 78,084 \\  \bottomrule
\end{tabular}
\caption{The sizes of the used data sets.}
\label{tab:data sets}
\end{wraptable} 

Similar supervised deep learning approaches were successful in discriminating between individual PSs, originating from different types of particle recoils \cite{holl_deep_2019,Khosa_2020,9747313}, or recoils in different positions and components of the detector \cite{AZCRESSTPulseShapeDiscr,Delaquis_2018,CMCRESSTPulseShapeDiscr,FWCRESSTPulseShapeDiscr}. 
The discrimination and reconstruction of pile-up artifacts was studied in Refs. \cite{fantini_machine_2022,https://doi.org/10.48550/arxiv.2112.06792}. The task of general data cleaning for cryogenic calorimeters with autoencoders (AEs) and variational AEs (VAEs) was studied in Refs. \cite{FWCRESSTPulseShapeDiscr,ichinohe_application_2022}, and with a Principal Component Analysis (PCA) in Ref.~\cite{Huang_2021}. Differently to those approaches, we do not rely on previous knowledge about the detector to which our method is applied, or the tuning of a cut value. We explore the synergies with the PCA method from Ref.~\cite{Huang_2021} in Sec.~\ref{sec:manifold}, by combining our approaches.
A method for the supervised discrimination between pulses and artifacts with neural networks was proposed in Ref. \cite{https://doi.org/10.48550/arxiv.2207.02187}, and shown on purely simulated data. We train and verify with both measured and simulated data, which adds necessary reliability to the results.



Our work was organized as follows: first, we create a large-statistics data set of labeled events. The procedure is described in Sec.~\ref{sec:data}. Second, we show that neural networks can learn the data cleaning task. The performance of our chosen models on the training set is presented in Sec.~\ref{sec:models}. Finally, we bridge the gap between triggering and the parts of the event selection which require detector-specific knowledge or tuning. In Sec.~\ref{sec:results}, we apply the trained models to a test set or recorded data and to simulated positive and negative events.



\section{Used data and pre-processing}
\label{sec:data}

The CRESST detector concept is based on a multi-channel read out. Within a joint housing (a detector module), a phonon and a light detector measure the produced phonon population and scintillation light from a particle recoil inside a scintillating crystal (the target). The detector response is measured with a Transition Edge Sensor (TES), operated in a read-out circuit with a SQUID amplifier. For a detailed description of TES we refer to Refs. \cite{seidel_phase_1990,probst_model_1995}. In some detector modules, additional TES serve to veto energy depositions in the housing or holding structure. In the context of this work, we treat all of them as individual detectors. 
We create our data set by extracting samples of several tens of hours measurement time from all measurements that were done between 2013 and 2019 in the CRESST setup at LNGS, for a total of 68 detectors and more than one million records. We exclusively use data from the $e^{-}/\gamma$ and neutron-calibration runs. These are dedicated measurement campaigns before or after the data for dark matter searches was taken. During the $e^{-}/\gamma$ calibration, a Co-57 source is used to produce a characteristic calibration line in the energy spectrum of the detector. During the neutron calibration, an AmBe source is used to produce a strong neutron flux and calibrate the response of the detectors for particles with no electromagnetic charge. By only using calibration data in this work we additionally prevent biasing effects for a potential application of the trained models to physics searches in future work. 
We chose seven detectors as a test set, these make up three detector modules: two times a target and one veto detector in a joint housing, and once a target and two veto detectors. These detectors were used only to evaluate the selection of the trained models, reported in \ref{sec:results}. For the optimization of hyperparameters of the classifiers and fit process, we split five percent of the data from the remaining 61 detectors into an individual validation set. The total sizes of the data sets are summarized in Tab.~\ref{tab:data sets}. The data preparation and cuts are done with the Python package `Cait' \cite{https://doi.org/10.48550/arxiv.2207.02187}. 

We want our classifier to successfully perform these operations:
\begin{itemize}
\item reject all jump, drift, spike, glitch and pile-up artifacts that deviate significantly from a recoil-type PS,
\item reject all PSs that rise far away from the trigger position at 1/4 of the record window or do not decay within the window, 
\item let all PSs survive that fit the above criteria, not only those from target recoils, and also if they show saturation effects typical for high energy recoils, 
\item let empty noise traces survive if their slope is within the typical slope of noise traces for the corresponding detector. 
\end{itemize}

\begin{wrapfigure}{r}{0.5\textwidth}
  \begin{center}
    \includegraphics[width=0.5\textwidth]{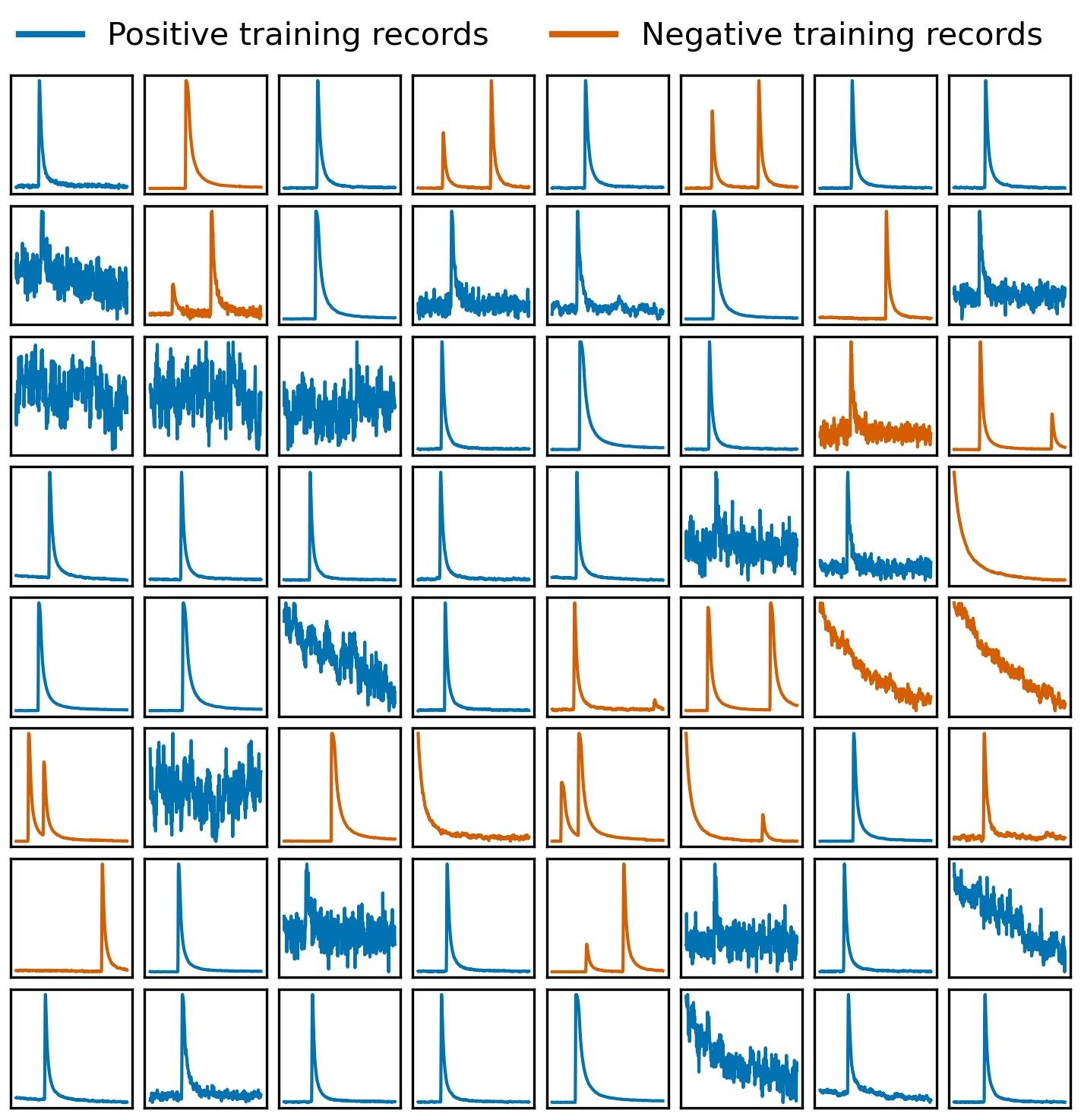}
  \end{center}
  \caption{A mini-batch of 41 positive (blue) and 23 negative (red) records from the training set, all from the same detector. About half of the negative records are created from positive ones, with a data augmentation technique (see text). At least one record (first row, second column) is wrongly labeled as negative.}
  \label{fig:training_samples}
\end{wrapfigure}

We apply to the triggered and noise events cuts on the PS parameters, to create the desired positive and negative labels for the classifier training. The used PS parameters include the PH, onset, rise and decay times of pulses, the difference between the offset levels on the left and right side of the record, the maximal and minimal derivatives and their positions within the record, and the mean, variance and skewness of the values in the record. We define rejection regions with the rectangle and lasso selection tools of Cait's VizTool, as described in Ref. \cite{https://doi.org/10.48550/arxiv.2207.02187}.

However, the applied cuts are imperfect, and generally tend to reject more pulses than necessary, which introduces a small share of wrongly labeled records in our data set. Their impact on the trained classifiers and reported performance metrics is studied in Sec.~\ref{sec:testset}. The training set consists to 83.6\% of positive records. We counter this imbalance with data augmentation, explained later in this section, and a weighted loss function, explained in Sec.~\ref{sec:models}.

We applied several transformations to the records as a preprocessing. The recordings were initially made with a sampling frequency of 25 kHz and record window lengths of 8192 or 16384 samples, depending on the measurement campaign (run). First, we downsampled the records to the length of 512 samples. The values in the downsampled time series correspond therefore to a time interval of 0.64 or 1.28 ms, while those in the original records corresponded to 40 mus. This time resolution and record length is sufficient to distinguish pulses from artefacts, and the step significantly reduces the necessary computing power for optimization. In a second step, we scale the values within all records individually, such that every record's minimal value is 0 and maximal value is 1. The information of the amplitude of individual records is, without further information about the corresponding detector, arbitrary and all useful information is contained in the shape of the time series.

The classifier optimization is an iterative process, for which the training set is split in so-called mini-batches of 64 records each. Three potential biases were identified in our data set, and their impact was mitigated by data augmentation methods: a) while we have an over-density of positive records, we have only a relatively small amount of pile-up artefacts. However, they are the class of artifacts that is most difficult to clean from the data with standard cuts. We therefore randomly pick several of the positive records in each mini-batch and superpose them with time-shifted copies of themselves, to artificially create pile-up events. The shifts are chosen such that the superposed pulse appears at a random position within the record window. We then changed their label from positive to negative. b) The data set contains many records with relatively high Signal-to-Noise Ratio (SNR), because most recorded pulses originate from calibration sources with typical energy depositions far above the detection threshold. A common method to robustify neural networks against small numerical deviations is to randomly add Gaussian noise to the inputs. We apply that method and by that artificially create records with lower SNR. c) The record window is built such that the trigger time is at 1/4 of its length. While large deviations in the pulse onset from this position are interpreted as artefacts and rejected, pulses with small deviations, of the order of milliseconds, should still be accepted. To implement this objective in our data set, we randomly shift all records by up to 26 samples (33.28 or 16.64 ms). The augmentations are applied to a record when it is drawn from the training set. Augmentation c) is always applied, but a) and b) only with a probability of 20\%. 

A mini-batch of 64 records from the training set is shown in Fig.~\ref{fig:training_samples}.

\section{Models and training process}
\label{sec:models}

A neural network classifier is a function $f_{w}$, which is parameterized with so-called weights $w \in \mathbb{R}^M$, that maps an element $x$ from a data set $\mathcal{D} \subset \mathbb{R}^N$ to a prediction $\hat{y} \in (0,1)$. A lower (higher) value indicates a higher belief that $x$ corresponds to a negative (positive) label $y \in \{0,1\}$. The number $N$ corresponds to the dimensionality of the data, namely the number of samples in the record (N=512). The number $M$ describes the number of weights in the neural network, and depends on the chosen model. Compared to other commonly used fit models, e.g. polynomials or splines, which typically have maximally several tens of parameters, neural networks have from thousands up to billions of parameters. In the limit $M \rightarrow \infty$, neural networks are proven to be universal function approximators \cite{hornik_multilayer_1989}. Furthermore, while the computational cost of many other function approximators increases exponentially with their number of parameters, neural networks do not suffer from this curse of dimensionality. These properties make them useful for the classification of high dimensional and strongly correlated data, which is the case for our raw sensor signals. For a high-quality pedagogical introduction to neural networks and machine learning we recommend Ref.~\cite{MEHTA20191}.

\begin{figure*}
 \includegraphics[width=\linewidth]{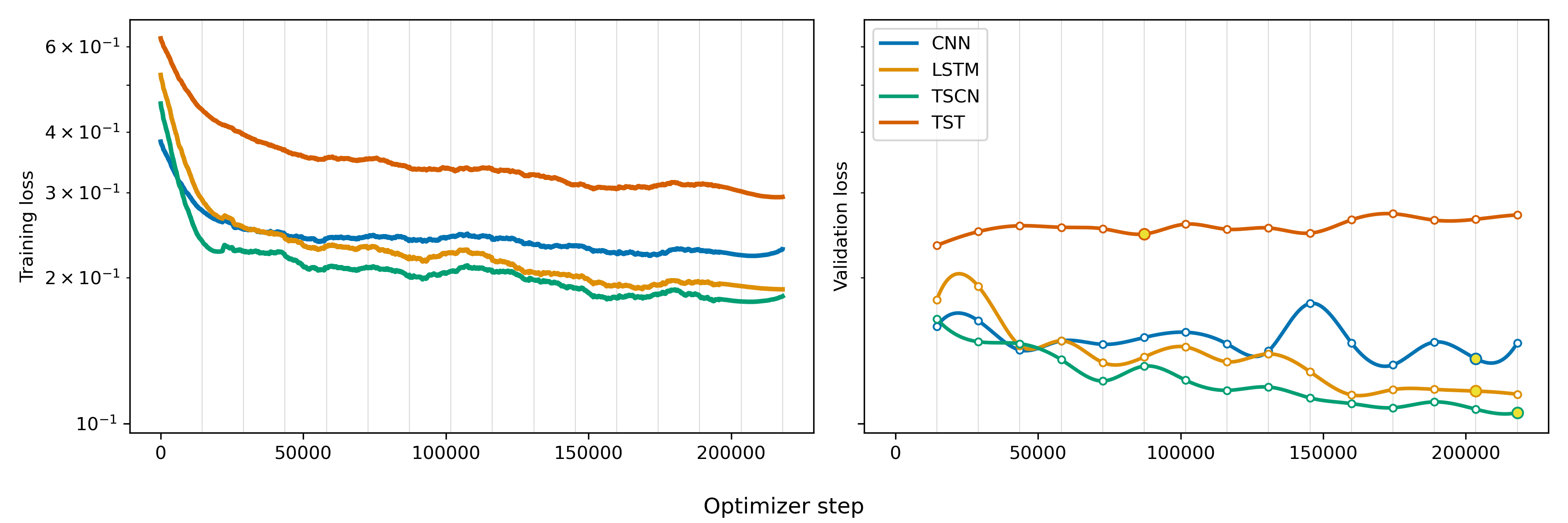}
\caption{Progression of loss values throughout the training process for the four considered models. (left) Loss on the training set, recorded for each optimizer step. (right) The loss on the validation set is evaluated at the end of each epoch. The spline interpolation is a guide for the eye. The yellow dots indicate the point in the training process, where the model reached the best agreement between labels and predictions (accuracy) on the validation set. The bumps in the validation loss, clearly visible for the CNN around 150k steps, are a typical artefact of stochastic optimizers.}
\label{fig:training_curves}       
\end{figure*}

For this work, we used four different models: 
\begin{itemize}
\item A small Convolutional Neural Network (CNN). This model applies a set of filters to the input, to extract meaningful features. It is highly efficient for spacially correlated data, e.g. for images. Our CNN applies successively 3 layers of filters, and extracts 64 feature channels. The filter kernels are weights, i.e. they are optimized jointly with the classifier parameters.
\item A larger convolutional model, which we will call Time Series Convolutional Network (TSCN). The performance of convolutional neural networks often depends on their number of weights, i.e. their size. To compare the capabilities of both a small and a larger model, we include this network with 8 layers of filters and 128 feature channels.
\item A bidirectional Long Short Term Memory (LSTM) network \cite{hochreiter_long_1997}. This model accumulates the information from a sequence of inputs into two internal states, one optimized to recognize long term dependencies, and one to recognize short term dependencies. It is widely used for sequential data, e.g. for natural language processing and time series. We use two bidirectional layers, and internal states with the dimensionality 200.
\item A for Time Series classification optimized Transformer (TST) \cite{vaswani_attention_2017}. Transformers are attention-based models, that recently have shown great performance on multiple types of data, among them time series applications \cite{https://doi.org/10.48550/arxiv.2010.02803}. They are very efficient in recognizing dependencies between different parts of sequences and the new state-of-the-art large language models \cite{https://doi.org/10.48550/arxiv.1810.04805}. We use the implementation of Ref.~\cite{mvts_transformer}.
\end{itemize}
The specifications and the hyperparameters of the models used in this work are given in in App. \ref{app:architectures}. All implementations are built in PyTorch \cite{NEURIPS2019_9015}.

The weights of neural networks are iteratively optimized (learned) by minimizing an error (loss) function between its predictions and the labels.
This process is called the `training' of the model. 
We minimize the binary cross entropy loss function, a quantification of the prediction error commonly used for classifier training (\cite{MEHTA20191}, p. 49). We weight the loss resulting from positive and negative records individually to counteract the preponderance of positive records. This way, both groups have the same impact in the optimization process. 
A sweep through the whole training data set, i.e. one iteration through all mini-batches, is called an epoch. 
We train all models for 15 epochs. We randomize the mini-batches with a technique called weak shuffling: individual mini-batches consist of records from the same detector, while the order in which mini-batches are used in the training process is randomized. This technique significantly reduces the data loading time during training compared to randomization within each mini-batch, as the records are stored consecutively on our hard drive. The optimization process was done on a Tesla P100 GPU and with the ADAptive Moment estimation (ADAM) optimizer \cite{https://doi.org/10.48550/arxiv.1412.6980}, a variant of stochastic gradient descent. The optimizer does one update of the weights (one optimizer step) for each mini-batch. The magnitude of weight updates in each optimizer step is steered by the learning rate. We optimized the learning rates for all models individually with the cyclic learning rate finder technique, described in Ref.~\cite{https://doi.org/10.48550/arxiv.1506.01186}. The learning rates used for each model are listed in Tab.~\ref{tab:learning rates}. 

After each training epoch, we evaluate the loss value on our validation set. The loss values on the training and validation set for all models throughout the training process are shown in Fig.~\ref{fig:training_curves}. All curves resemble the typical shape of neural network training curves, and in three out of four models the loss on the validation set continues to decrease proportionally to the loss on the training set. Only for the TST model, the validation loss rises slightly, which indicates overfitting on the training data. We experimented with the models hyperparameters to improve the overfitting, but other configurations lead to overall worse results on the validation set. We saved the trained model after each epoch, and can therefore use those which had the best agreement between labels and predictions (accuracy) on the validation set. The TSCN model shows the lowest loss value on the validation set after the training process. We do not apply the data augmentations (see Sec.~\ref{sec:data}) on the validation data, therefore the validation loss is generally lower than the training loss.

\section{Results}
\label{sec:results}

We run multiple tests with our trained models: we evaluate the balanced accuracy, recall and precision on the test set, the recall (selectivity) on simulated particle recoil (pile-up) events and we visualize the data manifold before and after application of our model with a PCA. Additionally, we compare a PH spectrum cleaned with classical cuts to one cleaned with our models. The metrics used in this section are defined as follows:

\begin{tabular}{llll}
    \text{Recall } &$R \coloneqq TP/T$, &\text{Selectivity } &$S \coloneqq TN/N$, \\
    \text{Balanced Accuracy } &$BA \coloneqq (R + S)/2$, &\text{Precision } &$P \coloneqq TP/(TP + FP)$, \\
    \text{Integral Over Recall } &$IOR \coloneqq \int_\Omega R(\mu) d\mu$, 
    &\text{Integral Over Selectivity } &$IOS \coloneqq \int_\Omega S(\mu) d\mu$, \\
\end{tabular}
\linebreak

\noindent where T are positive records, N are negative records, TP are true positive predictions and TN are true negative predictions. $R(\mu)$ and $S(\mu)$ are the recall and selectivity as functions of PS features. $\Omega$ is the region over which we integrate the recall or selectivity and $\mu$ is a placeholder for one or multiple PS features (e.g. the PH) w.r.t. which we perform the integration.

\begin{figure*}
 \includegraphics[width=.48\linewidth]{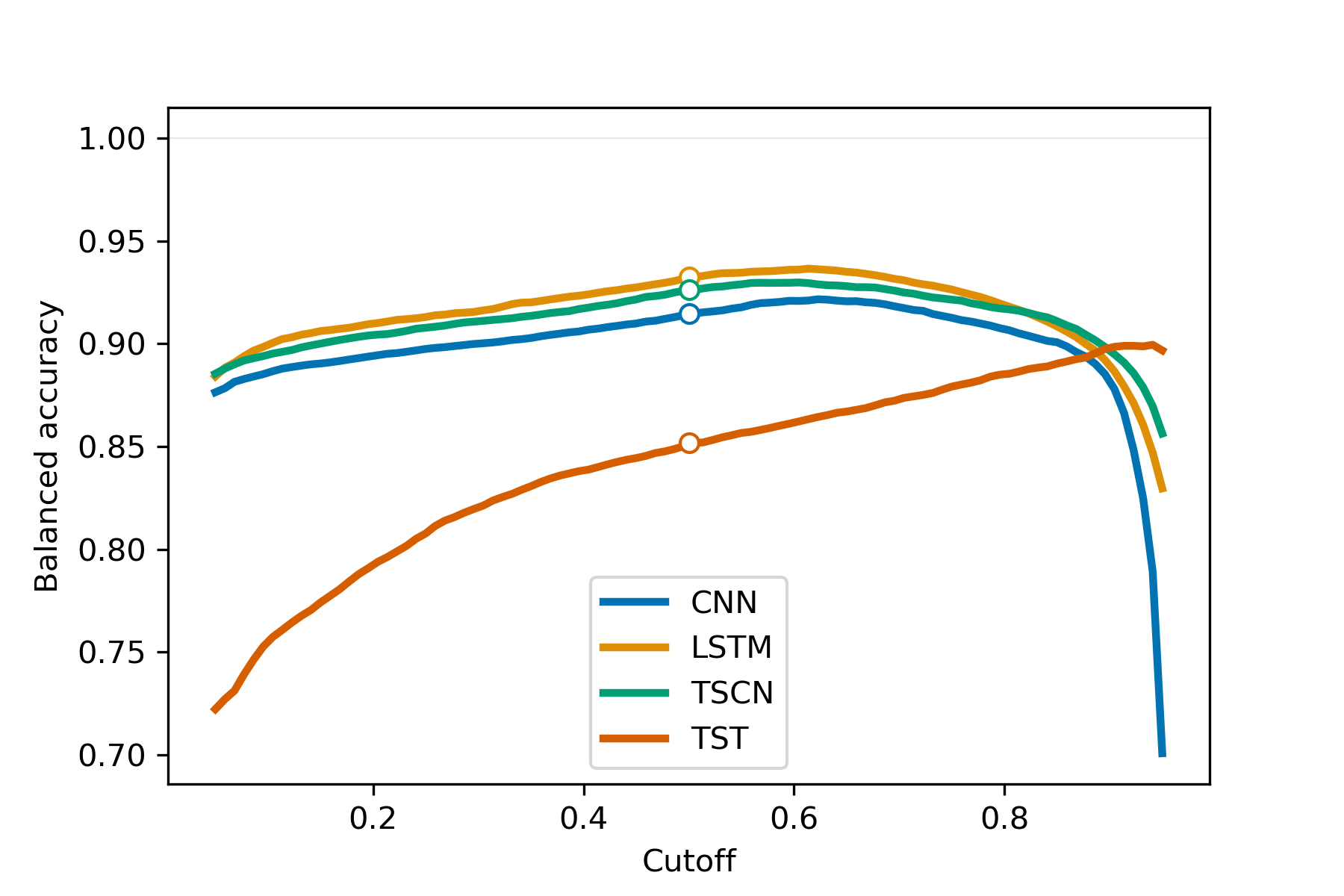}\hfill
 \includegraphics[width=.48\linewidth]{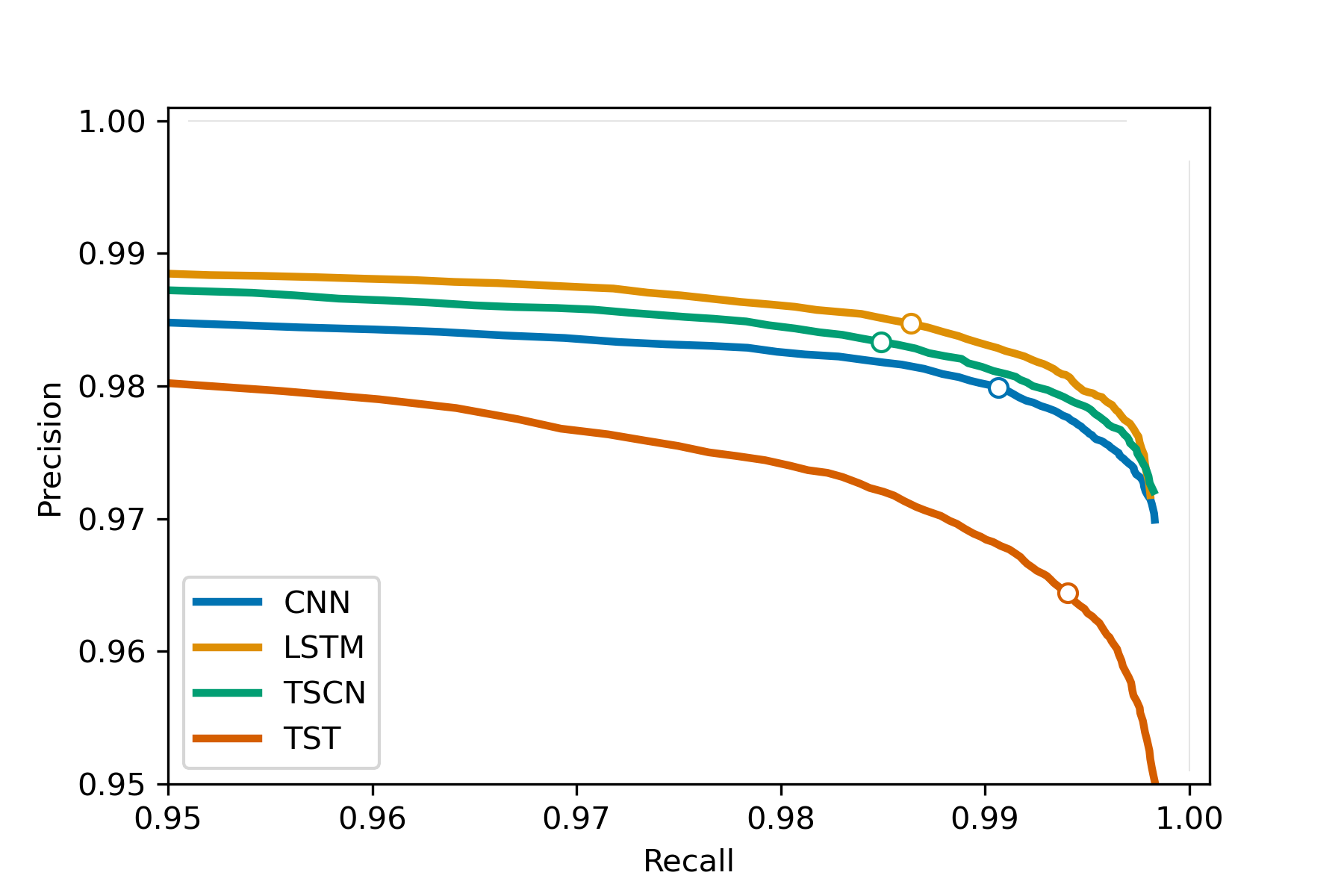}
\caption{Metrics of all classifier models, under varying cutoff values, evaluated on the test set. The white dot marks the default cutoff value of 0.5. (left) The balanced accuracy w.r.t.~the cutoff value. (right) The precision vs.~recall curves, for cutoff values between 0.05 and 0.95.}
\label{fig:metrics}       
\end{figure*}

\subsection{Evaluation on the test set}
\label{sec:testset}

\begin{wraptable}{r}{12cm}
\begin{tabular}{lcccccc}\\\toprule  
Model & Accuracy (b.) & Recall        & Precision     & IOR           & IOS           & Runtime (s)      \\\midrule
CNN   & .915          & .991          & .98           & .970          & .905          & \textbf{1.9 (1)} \\  \midrule
LSTM  & \textbf{.932} & .986 & \textbf{.985} & .993          & .97 & 12.5 (6.5)       \\  \midrule
TSCN  & .926          & .985          & .983          & .987          & \textbf{.976}         & 3.0 (1.6)        \\  \midrule
TST   & .852          & \textbf{.994}          & .964          & \textbf{.998} & .689          & 23.2 (12.1)      \\  \bottomrule
\end{tabular}
\caption{The metrics of the trained models, evaluated on the test set and simulated data. (col. 2-4) The balanced accuracy, recall and precision on the test set with a cutoff value of 0.5. (col. 5, 6) The IOR score for the simulated positive particle recoils, and the IOS score for the simulated negative pile-up events. The values are defined in the text. (col. 7) The runtime for predicting the whole test set. In brackets is the runtime divided by the lowest runtime.}
\label{wrap-tab:1}
\end{wraptable} 

We apply all models to the test set. To use the models as classifiers, a cutoff on the output value (the belief) has to be defined, below which we reject the record. We evaluate the balanced accuracy, the precision and the recall w.r.t. the cutoff value. They are visualized in Fig.~\ref{fig:metrics}. For specific applications, the cutoff value can be tuned to fulfill specific demands to recall or precision. In this work, we choose a generic cutoff value of 0.5 for all classifiers.  

The recall and precision are naturally in a opposing relationship, explaining the typical kink in the curve shape. The LSTM and TSCN models give the best performance across all metrics. The metrics for the cutoff value 0.5 are reported in Tab.~\ref{wrap-tab:1}. The predictions on the test set were done in batches as well, with a batch size of 32. The total run times for the predictions on the test set are reported in Tab.~\ref{wrap-tab:1}. There are typically several thousand predictions done per second, which is due to the low computational cost for inference with neural networks, and the strong parallelization of the necessary matrix multiplications on the GPU.

\begin{wrapfigure}{r}{0.5\textwidth}
  \begin{center}
    \raisebox{0pt}[\dimexpr\height-3\baselineskip\relax]{\includegraphics[width=0.48\textwidth]{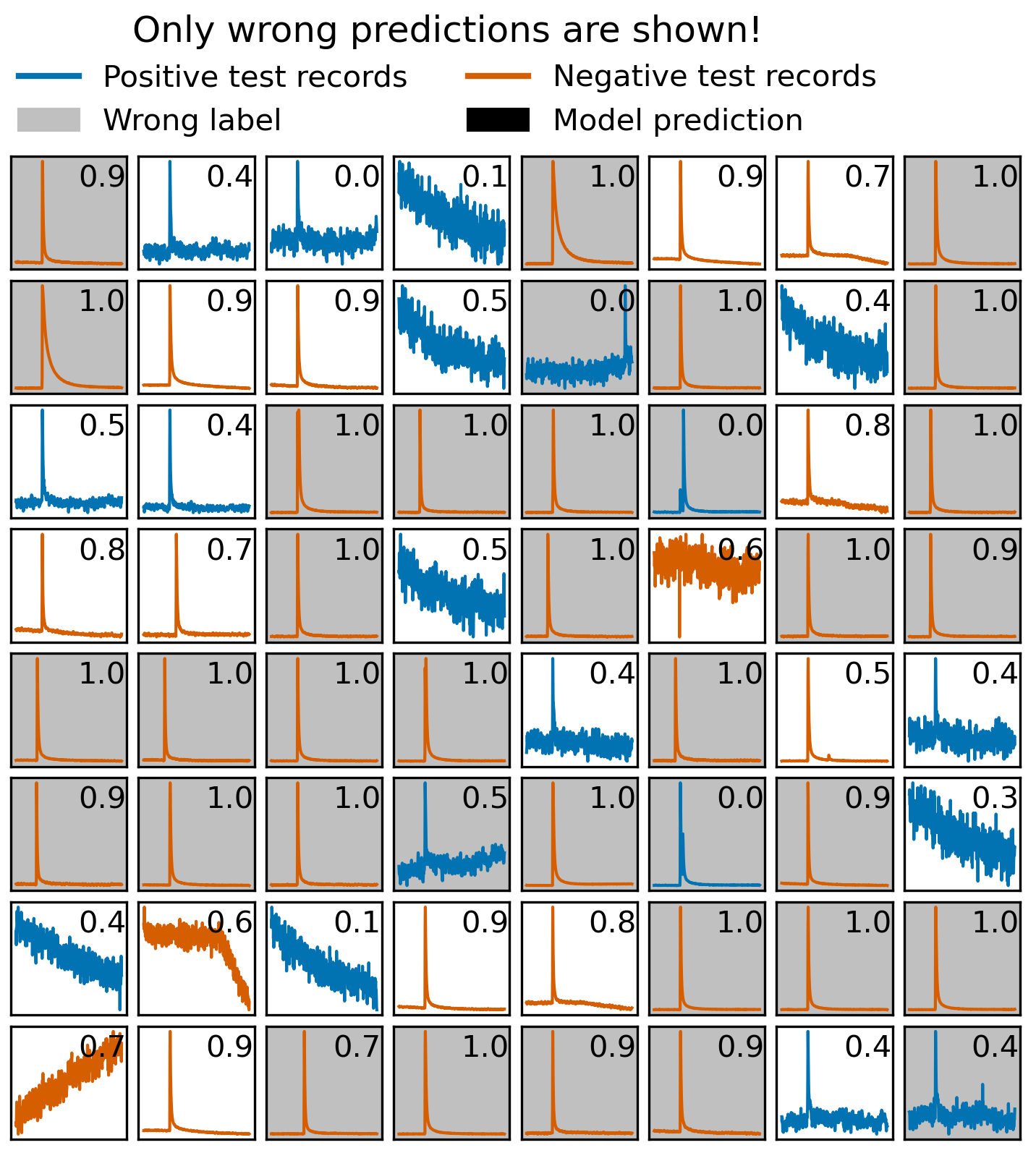}}%
  \end{center}
  \caption{A batch of events from the test set that were wrongly predicted by the LSTM. The grey color indicates wrong labels. Some records, among them the tilted BLs, can hardly be flagged as positive or negative without additional context, namely the distribution of the remaining data of the corresponding detector.}
  \label{fig:wrong_phonon}
\end{wrapfigure}

We take a closer look to the wrongly predicted records. For this, we use the predictions of the LSTM network, and pick one of the detectors from the test set. The detector has 70 records that were wrongly predicted, out of 8422 records in total. We show the first 64 of the wrongly predicted records in Fig.~\ref{fig:wrong_phonon}. After visual inspection, we discover that 39 of them are wrongly labeled, mostly good pulses that were collaterally rejected by the imperfect quality cuts on the main parameters. Among the wrongly labeled records there are also noise traces for which the decision if they should survive quality cuts or not depends on the distribution of noise traces in the individual detector. In a full analysis of a detector, these events would most likely not surpass the trigger threshold and are therefore irrelevant.

\subsection{Evaluation on simulated data}

\begin{figure*}
 \includegraphics[width=.48\linewidth]{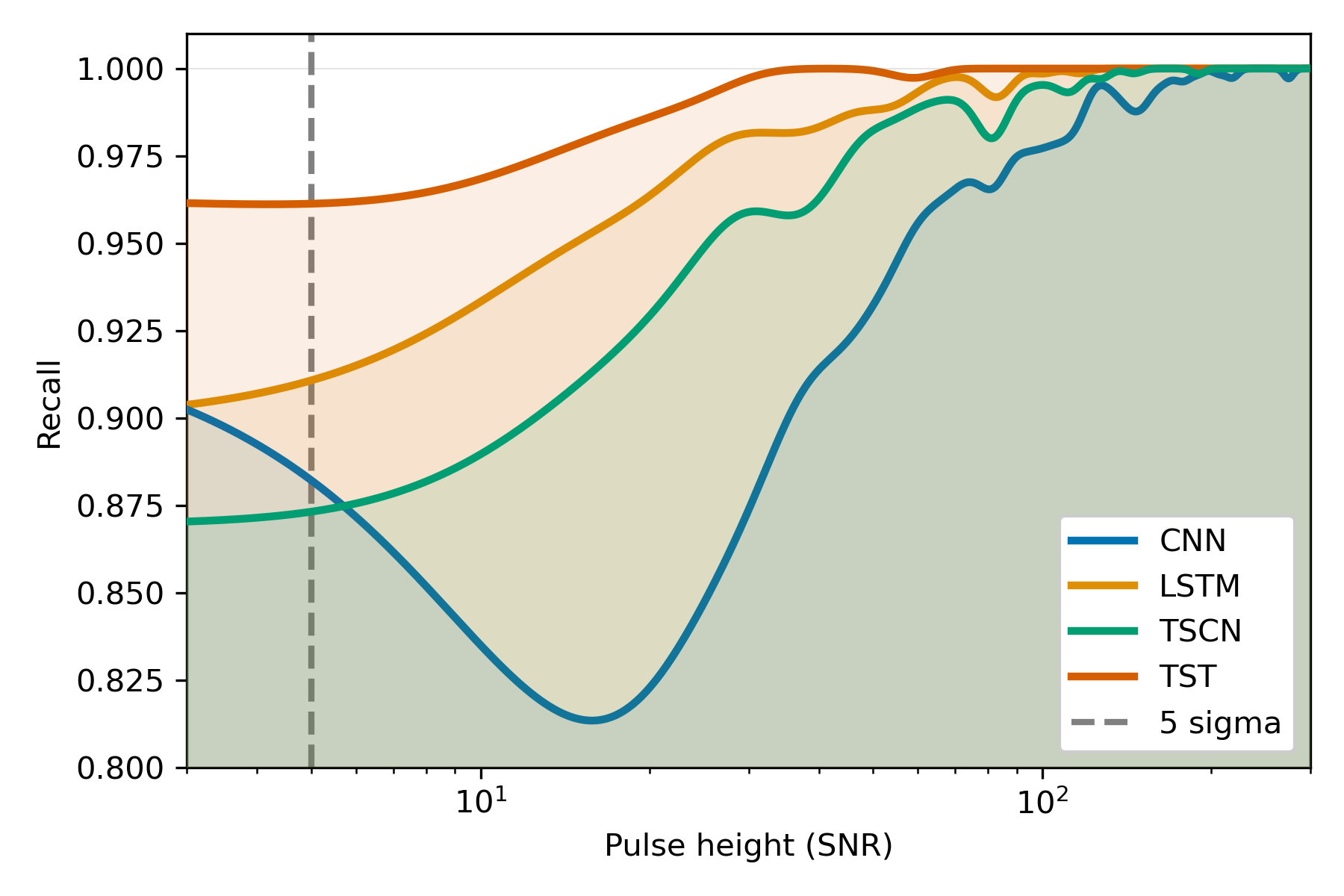}\hfill
 \includegraphics[width=.48\linewidth]{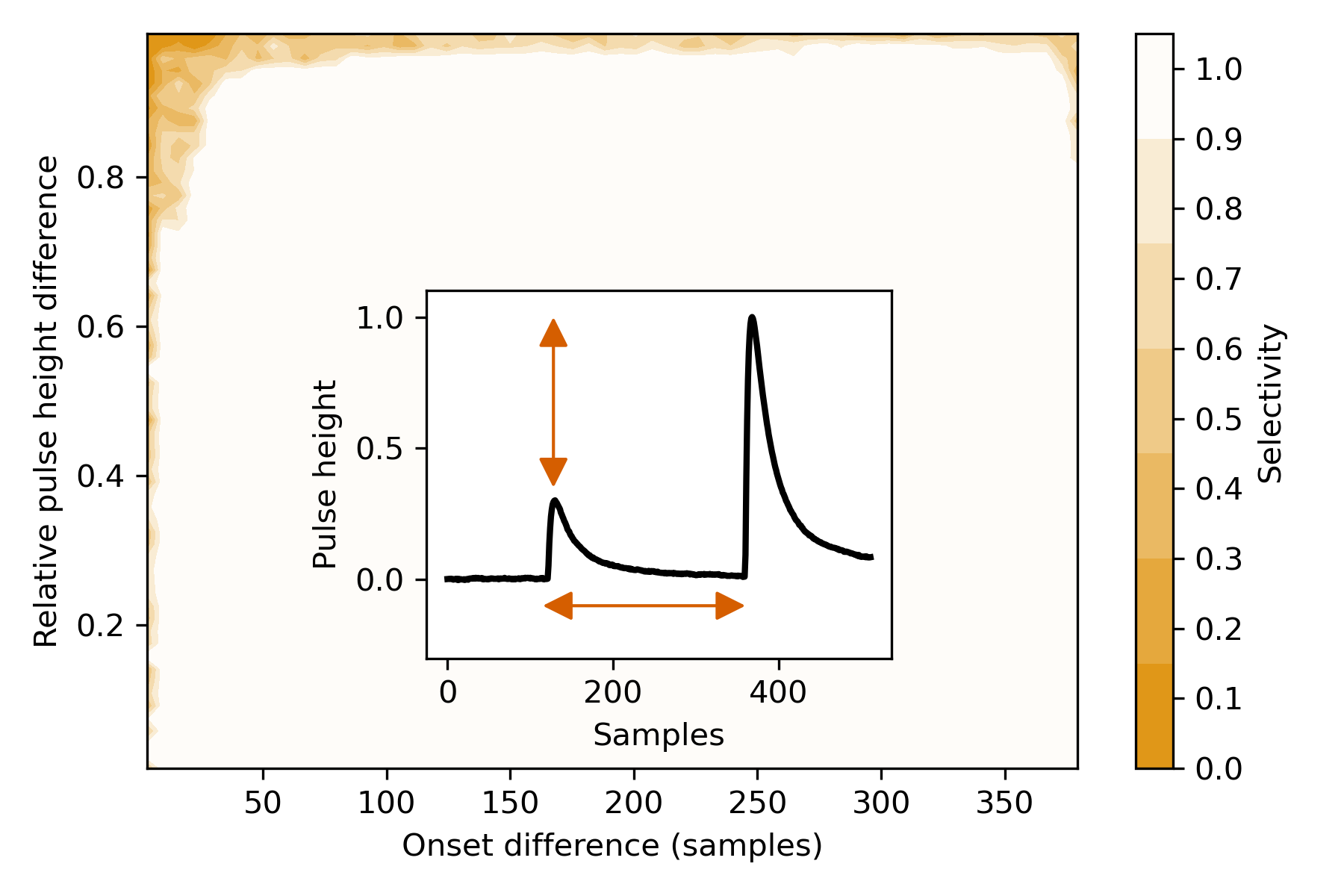}
\caption{Metrics of the classifier models, evaluated on simulated data. (left) The recall values w.r.t. the SNR of simulated events. The recall drops towards lower values, but is still reasonably high around a typical trigger threshold value of 5 BL noise resolutions (grey, dashed). The reason for the local minimum of the CNN curve above 10 SNR is not cogently clearified. The most likely hypothesis is the absence of many low energy pulses in the training set, which can introduce a bias in models predictions. The simultaneous dip in the recall of multiple models around 80 SNR is a small sample effect of the simulation: it could be connected to two simulated events with similar energy, with relatively strongly tilted BLs. (right) The selectivity values for the LSTM model on simulate pile-up events, featuring two pulses, w.r.t. the difference in onset and relative difference in PH. Only pile-up events with large relative PH difference or very small onset difference are not rejected by the model. The area that is covered by the inset holds only selectivity values of one. (right, inset) An example of a simulated pile-up event.}
\label{fig:simulated_metrics}       
\end{figure*}

The evaluation on data labeled by cuts has the drawback of wrong labels, we therefore do a second evaluation on simulated data. For this, we superpose the typical pulse shape of one detector from our test set onto empty noise traces that were generated according to Ref. \cite{CARRETTONI20101982}. We simulate a data set of 50000 positive particle recoil records and negative pile-up records each, with PHs between zero and three hundred SNR, i.e. the ratio of the PH of the event and the BL noise resolution. For the pile-up events the onset of the first pulse is placed to 1/4 of the window, the second is placed randomly within the window. We evaluate the recall on the positive records w.r.t.~the SNR. The result is visualized in Fig.~\ref{fig:simulated_metrics} (left). We calculate the IOR for the SNR-range from three to three hundred. Furthermore, we evaluate the selectivity on the pile-up events (Fig.~\ref{fig:simulated_metrics}, right), w.r.t the difference in onset and relative difference in PH. We calculate the IOS over the whole plotted range. Both IOR and IOS values are reported for all models in Tab.~\ref{wrap-tab:1}. The LSTM and TSCN score best and equally good in IOS, while the TST scores best in IOR. Later result is however due to the fact, that the TST defaults to high output values, which is also visible in Fig.~\ref{fig:metrics} (left). The convolutional models CNN and TSCN feature a significantly lower runtime as the LSTM and TST.

\subsection{Pulse height spectrum and data manifold}
\label{sec:manifold}

As a last evaluation of our model, we choose a 3-channel detector module from the test set. For this detector, the target is a cylindrical calcium tungstate crystal. The first channel of the module is a TES placed on the target, collecting the thermal and athermal phonons produced in particle recoils. The second channel is a TES placed on a shielding ring structure around the first TES, which is directly connected to both the target and the light detector. The third channel is a TES on the light detector, a beaker-shaped object that fully surrounds the target. 

\begin{figure*}
 \includegraphics[width=\linewidth]{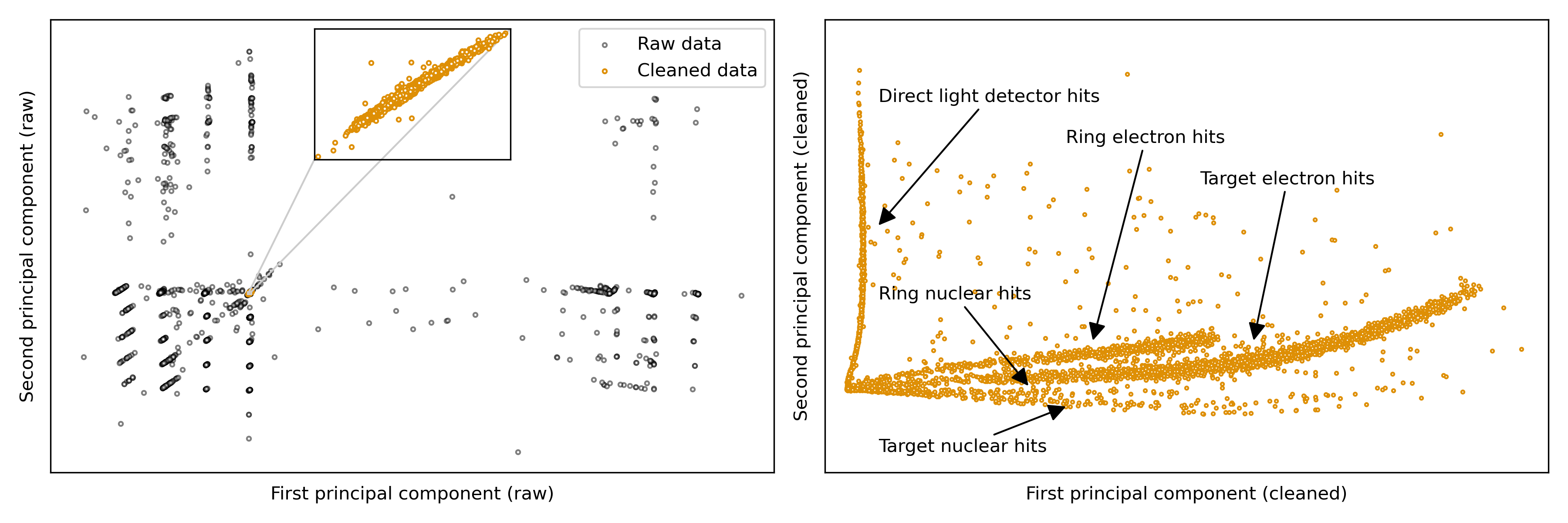}
\caption{The data manifold visualized with the first first two principal components. (left) The raw data, without cleaning (black) and the cleaned data (orange), both projected to the first and second principal components of the raw data matrix. (right) The cleaned data projected to the first and second principal components of the cleaned data matrix. The lines corresponding to the individual event types are clearly visible. The PH spectrum of the target channel is shown in Fig.~\ref{fig:spectrum}.}
\label{fig:pca}       
\end{figure*}

We apply our LSTM model to all channels of the detector module, and call the events clean that correspond to positive prediction for all three channels. In a technique called manifold visualization we want to visualize the distribution of the data in the high dimensional vector space of their sample values. To include the information of all three channels, we concatenate the time series of the records that correspond to the same event in different channels to vectors of length 1536, and form by that a combined data matrix. We do this for all data, and only the cleaned data, for which we use a PCA. The PCA is a singular value decomposition, i.e. the calculation of eigenvectors and eigenvalues, of the covariance matrix of the combined data matrix. The first (second) eigenvector of this matrix, called the first (second) principal component (PC), is itself a time series of length 1536 that corresponds to the template that accounts for the highest (second highest) variance of values in the concatenated records. For a more detailed description of the PCA method, we refer to Ref.~\cite{Huang_2021}, where it was first applied for pulse shape identification in cryogenic calorimeters. We plot the projection of the full raw data to its first and second PCs in Fig.~\ref{fig:pca} (left) and the same for only the cleaned data in Fig.~\ref{fig:pca} (right). The projection is calculated by building the dot product of the concatenated records and the PC. This corresponds to a change of basis in the high dimensional vector space of the sample values, with the first and second PCs as the new basis vectors. While the variance in the raw data is dominated by steps in the record windows induced by flux quantum losses of the SQUID amplifier (example in Fig.~\ref{fig:events}, orange), the cleaned data manifold resembles the actual particle recoil types present in the data: target hits, hits in the ring veto detector, and direct hits of the light detector. For the target and veto detector hits electron and nuclear recoils are also separately visible in the plot, due to the different amplitude in the light channel. The identification of the lines of individual event types in Fig.~\ref{fig:pca} (right) was done by comparing the equivalent plot for $e^-/\gamma$ and neutron calibration, and visual inspection of events and principal components individually. The first (second) principal component strongly resembles the PS of phonon-only (light-only) events. Thus, the line structure we observe in Fig.~\ref{fig:pca} (right) is in agreement with our expectation. Direct light detector hits, with no phonon signal, cluster along the second PC. Target hits cluster mostly along the first PC, but they differ in the amount of produced scintillation light. Electron hits produce significantly more scintillation light, and ring hits produce an additional signal in the light detector due to a weak thermal link between them.

\begin{wrapfigure}{r}{0.5\textwidth}
  \begin{center}
    \raisebox{0pt}[\dimexpr\height-1\baselineskip\relax]{\includegraphics[width=0.5\textwidth]{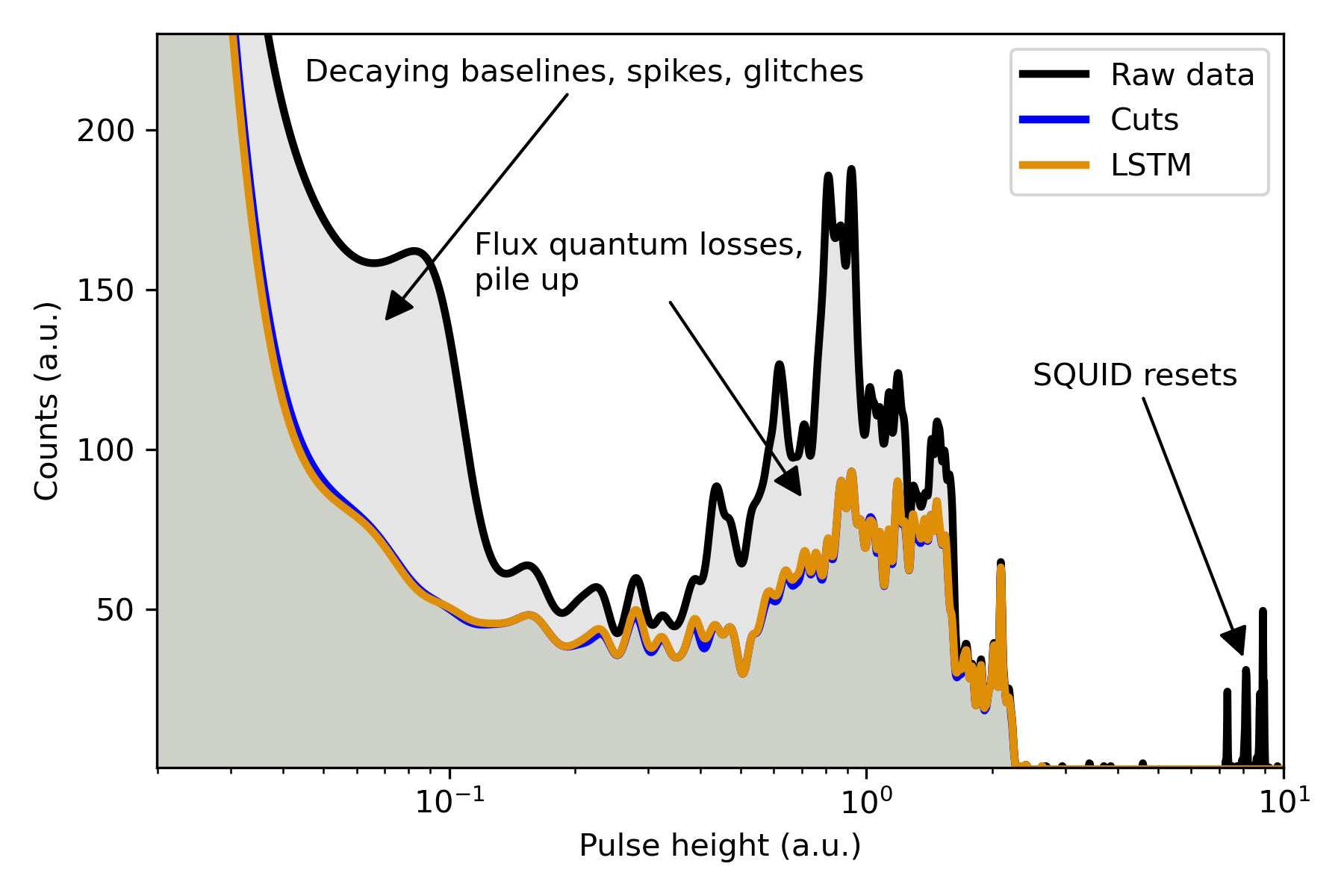}}%
  \end{center}
  \caption{The PH spectrum of an exemplary detector without cleaning (black), with the cut analysis that we used as labels (blue) and the LSTM predictions (LSTM). The blue and orange curves almost fully overlap due to the strong agreement between cuts and LSTM. The data manifold of the corresponding 3-channel detector module is visualized in Fig.~\ref{fig:pca}.}
  \label{fig:spectrum}
\end{wrapfigure}

Additionally, we show the PH spectrum of the target channel of the same detector module without any cuts, with the analysis cuts that we used as labels, and cutted with the LSTM model, in Fig.~\ref{fig:spectrum}. The PH is for low energy particle recoils proportional to the recoil energy. We observe a strong agreement of the LSTM model and the analysis cuts over the full width of the spectrum. The uncleaned data is strongly polluted with artifacts.

\vspace{.5cm}

To summarize, we have shown in Sec. \ref{sec:results}, that we reliably discriminate events originating from particle recoils and artefacts. Differently to standard approaches, we do not rely on prior knowledge of individual detectors, as its characteristic pulse shape, or manual interventions, as finding individual cut values. We have shown in Sec. \ref{sec:manifold}, that our trained models reproduce the objectives of the cuts that were made by analysts. The event selection, and with that the physics reach of the experiment, does therefore not change significantly whether our model is applied or the selection is done per hand. This is the outcome we had hoped for and our results can have the following impacts:
\begin{enumerate}
\item For large-scale detector setups the fine tuning of dedicated cuts for each detector can produce non-negligible delays in the analysis or even become infeasible. The application of our models instead can produce equivalent cuts instantly.
\item Recorded data can be monitored in real time. This can uncover unwanted shifts of the measurement setup, as e.g. an increased rate of artifacts, immediately and enable fast interventions. Furthermore, features in the event distribution, e.g. a peak in the PH spectrum, can be identified as particle-like or artefact-like, without the need for designing cuts first.
\end{enumerate}

\section{Conclusion and outlook}
\label{sec:conclusion}

We trained a selection of deep learning models to perform the data cleaning task for cryogenic detectors. We observed equally promising performance from a convolutional neural network, TSCN, and an LSTM neural network. The best achieved balanced accuracy score on the test set is 0.932, with a recall of 0.986 and precision of 0.985. Notably, the majority of the wrongly predicted records by the best performing model are either wrongly labeled records or hard-to-decide records. The method therefore apparently outperforms the cuts which we performed on the data, in the task of discriminating pulse-shaped from other records. For practical applications, our method has another advantage: the runtime is with O(1k-10k) predictions per second significantly faster than standard fit methods that are a typical alternative for the same purpose. Note, that our method can be applied to a new detector blindly, i.e. without relying on any information from the detector itself, as typical PSs. We therefore reduce the necessary manual task in data cleaning from handcrafting cut values for each detector individually, to merely controlling the predictions of the model. 

In future work, our method can be extended with an appropriate, unsupervised clustering of the different PSs. A promising approach was presented in Ref.~\cite{ichinohe_application_2022}. Furthermore, an investigation of the poor Transformer performance could be started. Likely the extraction of proper representations of the time series, as done for raw audio data in the popular wave2vec 2.0 framework \cite{10.5555/3495724.3496768}, would improve its performance. 

%
%
%

\section*{Acknowledgments}

We are grateful to LNGS for their generous support of CRESST. This work has been funded by the Deutsche Forschungsgemeinschaft (DFG, German Research Foundation) under Germany's Excellence Strategy – EXC 2094 – 390783311 and through the Sonderforschungsbereich (Collaborative Research Center) SFB1258 ‘Neutrinos and Dark Matter in Astro- and Particle Physics’, by the BMBF 05A20WO1 and 05A20VTA and by the Austrian Science Fund (FWF): I5420-N, W1252-N27 and FG1 and by the Austrian research promotion agency (FFG), project ML4CPD. The Bratislava group acknowledges a partial support provided by the Slovak Research and Development Agency (project APVV-15-0576). The computational results presented were obtained using the Vienna CLIP cluster.

\section*{Statements and Declarations}

On behalf of all authors, the corresponding author states that there is no conflict of interest. The data sets generated during and/or analysed during the current study are available from the corresponding author on reasonable request.

\clearpage
\newpage
\appendix

\section{Model architectures and learning rates}
\label{app:architectures}

The names of our tested architectures are chosen arbitrarily. The architecture of the CNN, LSTM and TSCN are is listed in Tab.~\ref{tab:cnnarch}, Tab.~\ref{tab:lstmarch} and Tab.~\ref{tab:tscnarch} respectively. The hyperparameters for the TST are listed in Tab.~\ref{tab:tstarch}, for a detailed list of the architecture we refer to Ref.~\cite{vaswani_attention_2017} and \cite{mvts_transformer}. The used learning rates are summarized in Tab.~\ref{tab:learning rates}.

\begin{table}
\centering
\begin{tabular}{ll}\\\toprule  
Layer&Specifications\\\midrule
1D convolutional& kernel size 8, 4 output channels, pooling 4, stride 1, ReLU activation\\\midrule
1D convolutional& kernel size 8, 16 output channels, pooling 4, stride 1, ReLU activation\\\midrule
1D convolutional& kernel size 8, 64 output channels, pooling 4, stride 1, ReLU activation\\\midrule
Fully connected& 320 input nodes, 200 output nodes, ReLU activation function\\\midrule
Fully connected& 200 input nodes, 84 output nodes, ReLU activation function\\\midrule
Fully connected& 84 input nodes, 1 output nodes, sigmoid activation function\\ \bottomrule
\end{tabular}
\caption{The details of the CNN architecture.}
\label{tab:cnnarch}
\end{table}

\begin{table}
\centering
\begin{tabular}{ll}\\\toprule  
Layer&Specifications\\\midrule
Bidirectional LSTM& 3 layers, input size 8, hidden size 200 \\\midrule
Fully connected& 12800 input nodes, 1 output node, sigmoid activation function\\ \bottomrule
\end{tabular}
\caption{The details of the LSTM architecture.}
\label{tab:lstmarch}
\end{table}

\begin{table}
\centering
\begin{tabular}{ll}\\\toprule  
Layer&Specifications\\\midrule
1D convolutional& kernel size 3, 16 output channels, stride 1, padding 1, ReLU activation\\\midrule
1D convolutional& kernel size 3, 16 output channels, pooling 2, stride 1, padding 1, ReLU activation\\\midrule
1D convolutional& kernel size 3, 32 output channels, stride 1, padding 1, ReLU activation\\\midrule
1D convolutional& kernel size 3, 32 output channels, pooling 2, stride 1, padding 1, ReLU activation\\\midrule
1D convolutional& kernel size 3, 64 output channels, stride 1, padding 1, ReLU activation\\\midrule
1D convolutional& kernel size 3, 64 output channels, pooling 2, stride 1, padding 1, ReLU activation\\\midrule
1D convolutional& kernel size 3, 128 output channels, stride 1, padding 1, ReLU activation\\\midrule
1D convolutional& kernel size 3, 128 output channels, pooling 2, stride 1, padding 1, ReLU activation\\\midrule
Fully connected& 4096 input nodes, 1024 output nodes, ReLU activation function\\\midrule
Fully connected& 1024 input nodes, 512 output nodes, ReLU activation function\\\midrule
Fully connected& 512 input nodes, 1 output nodes, sigmoid activation function\\ \bottomrule
\end{tabular}
\caption{The details of the TSCN architecture.}
\label{tab:tscnarch}
\end{table}

\begin{table}
\centering
\begin{tabular}{ll}\\\toprule  
Hyperparameter& Value\\\midrule
n.blocks& 1\\ \midrule
n.heads& 8\\ \midrule
dim. model& 64\\ \midrule
dim. FFW& 256\\ \bottomrule
\end{tabular}
\caption{The hyperparameters of the TST architecture.}
\label{tab:tstarch}
\end{table}

\begin{table}
\centering
\begin{tabular}{lr}\\\toprule  
Model & Learning rate \\\midrule
CNN & $5 \cdot 10^{-4}$ \\  \midrule
LSTM & $10^{-4}$ \\  \midrule
TSCN & $5 \cdot 10^{-5}$ \\  \midrule
TST & $10^{-5}$ \\  \bottomrule
\end{tabular}
\caption{The learning rates used for the training process of the models.}
\label{tab:learning rates}
\end{table} 

%
\bibliographystyle{utphys}
\bibliography{main}
%
%
%

\end{document}